\documentclass [11pt,a4paper] {article}

\usepackage[cp1252]{inputenc}
\usepackage{amssymb}
\usepackage{amsmath}
\usepackage{amsfonts,amssymb}
\usepackage[dvips]{graphicx}
\usepackage{bbm}
\usepackage{enumerate}
\usepackage{amsthm}
\usepackage{cancel}

\DeclareMathAlphabet{\mathpzc}{OT1}{pzc}{m}{it}

\def\SmallColSep{\setlength{\arraycolsep}{1pt}}

\begin{document}

\title{Discrimination against or in favor of qubits in quantum theory}

\author{Arkady Bolotin\footnote{$Email: arkadyv@bgu.ac.il$\vspace{5pt}} \\ \emph{Ben-Gurion University of the Negev, Beersheba (Israel)}}

\maketitle

\begin{abstract}\noindent Within context of quantum logic, it is possible to assign dispersion-free probabilities to experimental propositions pertaining to qubits. This makes qubits distinct from the rest of quantum systems since the latter do not admit probabilities having only values 0 and 1. The present paper shows that erasing qubit discrimination leads to a model of computation which permits execution of many primitive operations in a massive parallel way. In the paper, it is demonstrated that such a model (that can be called a quantum parallel random-access machine, QPRAM) is quantum mechanically plausible.\\

\noindent \textbf{Keywords:} Truth value assignment; Experimental quantum propositions; Propositional semantic; Closed linear subspaces; Probability measure; Systems of linear equations; Solvability; Cost of computation; Quantum parallel computing.\\
\end{abstract}

\section{Introduction}  

\noindent A quantum bit (qubit for short) is a two-level quantum system, i.e., one whose corresponding Hilbert space is two-dimensional. It is the simplest quantum mechanical system that can exist. Therefore, it is reasonable to expect that – otherwise than the least complicated mathematical framework required for the analysis – there is no distinction against or in favor of a qubit in quantum theory. But, surprisingly, attempts to prove this apparently right claim using the mathematical formalism of quantum mechanics run into difficult questions underlying quantum logic and computer science.\\

\noindent In particular, within the scope of quantum logic, i.e., an orthomodular partially ordered set of closed linear subspaces of a Hilbert space $\mathcal{H}$ (as well as corresponding projection operators), qubits – unlike many-level quantum systems – cannot be included in the domain of Gleason’s theorem \cite{Gleason} which gives a motivation for a continuous probability measure on the abovementioned set. In this way, one of the fundamental theorems in quantum mechanics discriminates against qubits.\\

\noindent To eliminate the said discrimination, one may consider positive operators (so-called effects) instead of projection operators and define a probability measure as a function on those objects  \cite{Busch97, Busch03}. However, such a line of attack invalidates the natural link between closed linear subspaces of a Hilbert space and experimental quantum propositions. Meanwhile, the search for the alternative assumption(s) needed to include qubits into a Gleason-type theorem is still going on (as an example, see the recent papers \cite{Benavoli} and \cite{Wright}).\\

\noindent What is more, qubit discrimination goes further than Gleason's theorem.\\

\noindent Indeed, as it is known from elementary linear algebra (see, for example, \cite{Anton}), the closed linear subspace, say $\mathcal{P}$, can be mathematically represented using the subspace spanned by linearly independent column vectors of the matrix $\mathbf{M}(\hat{P})$ that encodes – with respect to some arbitrary orthonormal basis – the projection operator $\hat{P}$ accordant to the experimental quantum proposition $P$. Consequently, the statement declaring that some unit vector in $\mathcal{H}$, say $|\Psi\rangle$, belongs to $\mathcal{P}$ is equivalent to the statement that the matrix $\mathbf{\Psi}$ encoding $|\Psi\rangle$ (with respect to the chosen basis) is within the said span. That is, the statement $|\Psi\rangle\in\mathcal{P}$ is true together with the statement asserting the solvability of the linear system $\mathbf{RX}\mkern-3mu=\mkern-2mu\mathbf{\Psi}$, where $\mathbf{R}$ is the matrix that comprises the linearly independent column vectors of $\mathbf{M}(\hat{P})$.\\

\noindent On the other hand, if the statement $|\Psi\rangle\in\mathcal{P}$ is true, one can say that $|\Psi\rangle$ assigns the truth value of true to the experimental quantum proposition $P$ represented by the subspace $\mathcal{P}$. In this way, the truth assignment can be treated as a computational problem, namely, the inspection of solvability of the linear system $\mathbf{RX}\mkern-3mu=\mkern-2mu\mathbf{\Psi}$. Likewise, the assignment of the falsity to the proposition $P$ can be seen as the inspection of solvability of the linear system $\mathbf{KX}\mkern-3mu=\mkern-2mu\mathbf{\Psi}$, where $\mathbf{K}$ is the matrix that consists of the linearly independent column vectors of $\mathbf{I}-\mathbf{M}(\hat{P})$, the difference between the identical matrix $\mathbf{I}$ and the matrix $\mathbf{M}(\hat{P})$.\\

\noindent A puzzling matter is that the linear system $\mathbf{RX}\mkern-3mu=\mkern-2mu\mathbf{\Psi}$ has only one unknown, while the linear system $\mathbf{KX}\mkern-3mu=\mkern-2mu\mathbf{\Psi}$ has $n-1$ unknowns. This implies that for the case of $n>2$, i.e., for the experimental atomic propositions relating to many-level quantum systems, the number of primitive operations needed to assign the truth value of false must be greater than that of true. In contrast, for qubits, the truth and the falsity of experimental atomic propositions are expected to be equal in numbers of primitive operations. One can infer from this that the truth value assignment discriminates in favor of qubits.\\

\noindent The present paper offers a way to resolve the discrimination both against and in favor of qubits thus formally proving that qubits are part of the quantum realm.\\

\section{The problem of an extra object}  

\noindent Let a quantum system be associated with a (separable) Hilbert space $\mathcal{H}$, whose unitary vectors correspond to possible pure states of the quantum system. In accordance with Birkhoff and von Neumann's proposal \cite{Birkhoff}, the mathematical representative of some experimental proposition $P$ pertaining to the quantum system is $\mathrm{ran}(\hat{P})$, the range of the projection operator $\hat{P}$ accordant to the proposition $P$:\smallskip

\begin{equation} \label{RAN} 
   \mathrm{ran}(\hat{P})
   =
   \left\{
      |\varphi\rangle
      \in
      \mathcal{H}
      \textnormal{:}
      \mkern10mu
      \hat{P}
      |\varphi\rangle
      =
      |\varphi\rangle
      \mkern-2mu    
   \right\}
   \;\;\;\;  .
\end{equation}
\smallskip

\noindent By the same token, the mathematical representative of ${\neg}P$, the negation of the proposition $P$, is $\mathrm{ker}(\hat{P})$, i.e., the kernel of the projection operator $\hat{P}$:\smallskip

\begin{equation}  
   \mathrm{ker}(\hat{P})
   =
   \left(
      \mathrm{ran}(\hat{P})
   \right)^{\perp}
   =
   \mathrm{ran}(\hat{1} - \hat{P})
   =
   \left\{
      |\varphi\rangle
      \in
      \mathcal{H}
      \textnormal{:}
      \mkern10mu
      \hat{P}
      |\varphi\rangle
      =
      0
      \mkern-2mu    
   \right\}
   \;\;\;\;  ,
\end{equation}
\smallskip

\noindent where $\perp$ stands for the orthogonal complement property and $\hat{1}$ denotes the identity operator. Suppose that the system is in the pure state corresponding to the unit vector $|\psi\rangle$ which belongs to either $\mathrm{ran}(\hat{P})$ or $\mathrm{ker}(\hat{P})$. Then, one can say that $|\psi\rangle$ assigns the truth value of either true or false, respectively, to the proposition $P$.\\

\noindent This suggests that the truth value of any experimental atomic proposition in an arbitrary pure state can be determined by the agency of the relation ``is an element of'', $\in$ , between the unit vector $|\Psi\rangle$ describing the given state and two closed linear subspaces, $\mathrm{ran}(\hat{P})$ and $\mathrm{ker}(\hat{P})$, representing the given proposition, that is, $|\Psi\rangle\mkern-2.5mu\in\mkern-2mu\mathrm{ran}(\hat{P})$ and $|\Psi\rangle\mkern-2.5mu\in\mkern-2mu\mathrm{ker}(\hat{P})$.\\

\noindent To make this suggestion precise, let us use the double-bracket notation $[\mkern-3.3mu[\cdot]\mkern-3.3mu]$ to express a truth value of a mathematical statement or an experimental proposition (as well as a propositional formula constructed from atomic statements or experimental propositions).\\

\noindent Let $x$ be a mathematical statement, for example, $|\Psi\rangle\mkern-2.5mu\in\mkern-2mu\mathrm{ran}(\hat{P})$. Since $x$ is either true or false (but not both), the truth value of $x$ can be regarded as the image of $x$ under the function $v$, namely,\smallskip

\begin{equation}  
   v
   \mkern-3.3mu
   :
   \mkern2mu
   \mathbb{S}
   \to
   \mathbb{B}_{2}
   \;\;\;\;  ,
\end{equation}
\smallskip

\noindent where $\mathbb{S}$ is the set of mathematical statements and $\mathbb{B}_{2}$ is the set of two truth values, true and false (or 1 and 0). The image of $x$ under $v$ can be denoted ${[\mkern-3.3mu[x]\mkern-3.3mu]}_v = v(x)$.\\

\noindent Now, consider the map\smallskip

\begin{equation} \label{MAP} 
   b
   \mkern-3.3mu
   :
   \mkern2mu
   \mathbb{B}_{2}
   \times
   \mathbb{B}_{2}
   \to
   \mathbb{B}_{2}
   \;\;\;\;  ,
\end{equation}
\smallskip

\noindent where $b$ is the Boolean function\smallskip

\begin{equation}  
   {[\mkern-3.3mu[P]\mkern-3.3mu]}_b
   =
   b
   \left(
      {[\mkern-3.3mu[x]\mkern-3.3mu]}_v
      ,
      {[\mkern-3.3mu[y]\mkern-3.3mu]}_v
   \right)
   \;\;\;\;   
\end{equation}
\smallskip

\noindent that takes the truth values of two mathematical statements $x$ and $y$, which cannot be true together in any physically meaningful state, $|\Psi\rangle\neq0$, namely,\smallskip

\begin{equation}  
   x
   \mkern-3.3mu
   :
   \mkern5mu
   |\Psi\rangle
   \mkern-2.5mu
   \in
   \mkern-2mu
   \mathrm{ran}(\hat{P})
   \;\;\;\;  ,
\end{equation}
\\[-36pt]

\begin{equation}  
   y
   \mkern-3.3mu
   :
   \mkern5mu
   |\Psi\rangle
   \mkern-2.5mu
   \in
   \mkern-2mu
   \mathrm{ker}(\hat{P})
   \;\;\;\;  ,
\end{equation}
\smallskip

\noindent to elements of $\mathbb{B}_{2}$ and, by doing so, assigns the truth value to the experimental atomic proposition $P$. In particular, $b$ returns 1 if $x$ is true and $b$ returns 0 if $y$ is true. Symbolically, $b(1,0)=1$ and $b(0,1)=0$.\\

\noindent However, since the statement $|\Psi\rangle\mkern-2.5mu\notin\mkern-2mu\mathrm{ran}(\hat{P})$ is not logically equivalent to the statement $|\Psi\rangle\mkern-2.5mu\in\mkern-2mu\mathrm{ker}(\hat{P})$ (in symbols, ${\neg}x{\mkern2.5mu\nLeftrightarrow\mkern2mu}y$), it may be the case that the statements $x$ and $y$ are false together. Given three different objects, that is, three ordered pairs $(1,0)$, $(0,1)$ and $(0,0)$, but only two categories to put them into, i.e., the truth values of true and false, one has a problem (which can be called \emph{the problem of an extra object}): What is the image of the pair $(0,0)$ under the Boolean function $b$? In other words, what truth value should be assigned to the experimental atomic proposition $P$ in case the vector $|\Psi\rangle$ is a linear combination (superposition) of vectors in the subspaces $\mathrm{ran}(\hat{P})$ and $\mathrm{ker}(\hat{P})$? This problem can be symbolically presented as follows:\smallskip

\begin{equation} \label{PR1} 
   {[\mkern-3.3mu[P]\mkern-3.3mu]}_b
   =
   b
   \left(
      {[\mkern-3.3mu[x]\mkern-3.3mu]}_v
      ,
      {[\mkern-3.3mu[y]\mkern-3.3mu]}_v
   \right)
   =
   \left\{
      \begingroup\SmallColSep
      \begin{array}{r l}
         1
         &
         \mkern3mu
         ,
         \mkern12mu
         {[\mkern-3.3mu[x]\mkern-3.3mu]}_v
         =
         1
         \\[5pt]
         0
         &
         \mkern3mu
         ,
         \mkern12mu
         {[\mkern-3.3mu[y]\mkern-3.3mu]}_v
         =
         1
         \\[5pt]
         ?
         &
         \mkern3mu
         ,
         \mkern12mu
         {[\mkern-3.3mu[x]\mkern-3.3mu]}_v
         =
         {[\mkern-3.3mu[y]\mkern-3.3mu]}_v
         =
         0
      \end{array}
      \endgroup   
   \right.
   \;\;\;\;  .
\end{equation}
\smallskip

\noindent Furthermore, let $L(\mathcal{H})$ be a set of closed linear subspaces of a Hilbert space $\mathcal{H}$. Each two-element subset in $L(\mathcal{H})$, that is,\smallskip

\begin{equation}  
   \mathcal{S}(\hat{P})
   =
   \left\{
      \mathrm{ran}(\hat{P})
      \mkern2mu
      ,
      \mkern1mu
      \mathrm{ker}(\hat{P})
   \right\}
   \;\;\;\;  ,
\end{equation}
\smallskip

\noindent corresponds to a certain projection operator $\hat{P}$. Let two projection operators $\hat{Q}$ and $\hat{P}$ be distinct from each other and 0. The subsets $\mathcal{S}(\hat{Q})$ and $\mathcal{S}(\hat{P})$ can be called \emph{comparable} if their elements $\mathrm{ran}(\hat{Q})$ and $\mathrm{ran}(\hat{P})$ can be ordered by set inclusion (or reverse set inclusion), that is, if the logical disjunction $z$\smallskip

\begin{equation}  
   z
   =
   z_1
   \sqcup
   z_2
   \;\;\;\;   
\end{equation}
\smallskip

\noindent whose operands are the mathematical statements\smallskip

\begin{equation}  
   z_1
   \mkern-3.3mu
   :
   \mkern5mu
   \mathrm{ran}(\hat{Q})
   \mkern-2.5mu
   \subseteq
   \mkern-2mu
   \mathrm{ran}(\hat{P})
   \;\;\;\;  ,
\end{equation}
\\[-36pt]

\begin{equation}  
   z_2
   \mkern-3.3mu
   :
   \mkern5mu
   \mathrm{ran}(\hat{P})
   \mkern-2.5mu
   \subseteq
   \mkern-2mu
   \mathrm{ran}(\hat{Q})
   \;\;\;\;  ,
\end{equation}
\smallskip

\noindent where $\subseteq$ denotes set inclusion, is true. Note that when ${[\mkern-3.3mu[z]\mkern-3.3mu]}_v=1$, the projection operators $\hat{Q}$ and $\hat{P}$ commute (are compatible), namely, $\hat{Q}\hat{P}=\hat{P}\hat{Q}$.\\

\noindent Contrastively, the subsets $\mathcal{S}(\hat{Q})$ and $\mathcal{S}(\hat{P})$ can be called \emph{incomparable} if their elements $\mathrm{ran}(\hat{Q})$ and $\mathrm{ran}(\hat{P})$ are orthogonal to each other, that is, if the logical disjunction $w$\smallskip

\begin{equation}  
   w
   =
   w_1
   \sqcup
   w_2
   \;\;\;\;   
\end{equation}
\smallskip

\noindent whose operands are the mathematical statements\smallskip

\begin{equation}  
   w_1
   \mkern-3.3mu
   :
   \mkern5mu
   \mathrm{ran}(\hat{Q})
   \mkern-2.5mu
   \subseteq
   \mkern-2mu
   \mathrm{ker}(\hat{P})
   \;\;\;\;  ,
\end{equation}
\\[-36pt]

\begin{equation}  
   w_2
   \mkern-3.3mu
   :
   \mkern5mu
   \mathrm{ran}(\hat{P})
   \mkern-2.5mu
   \subseteq
   \mkern-2mu
   \mathrm{ker}(\hat{Q})
   \;\;\;\;  ,
\end{equation}
\smallskip

\noindent is true. Note that in case ${[\mkern-3.3mu[w]\mkern-3.3mu]}_v=1$, the projection operators $\hat{Q}$ and $\hat{P}$ are compatible and orthogonal, specifically, $\hat{Q}\hat{P}=\hat{P}\hat{Q}=0$.\\

\noindent Let the proposition asserting that the subset $\mathcal{S}(\hat{Q})$ is comparable to the subset $\mathcal{S}(\hat{P})$ be denoted by $\mathcal{S}(\hat{Q})\mkern-1.5mu\lesseqgtr\mkern-1mu\mathcal{S}(\hat{P})$. The truth value of this proposition can be determined using the map (\ref{MAP}) such that\smallskip

\begin{equation}  
   {[\mkern-3.3mu[\mathcal{S}(\hat{Q})\mkern-1.5mu\lesseqgtr\mkern-1mu\mathcal{S}(\hat{P})]\mkern-3.3mu]}_b
   =
   b
   \left(
      {[\mkern-3.3mu[z]\mkern-3.3mu]}_v
      ,
      {[\mkern-3.3mu[w]\mkern-3.3mu]}_v
   \right)
   \;\;\;\;  ,
\end{equation}
\smallskip

\noindent where $b$ is the Boolean function that takes truth values of two mathematical statements $z$ and $w$, which cannot be true together in any physically meaningful state, to elements of $\mathbb{B}_2$. Expressly, $b$ returns 1 if $z$ is true, and $b$ returns 0 if $w$ is true.\\

\noindent Again, like in the situation with an experimental atomic proposition, the problem of an extra object arises. To be sure, because ${\neg}z{\mkern2.5mu\nLeftrightarrow\mkern2mu}w$, the statement $z$ asserting ordering by set inclusion and the statement $w$ asserting orthogonality may be false together. So, given three ordered pairs $(1,0)$, $(0,1)$ and $(0,0)$, but only two categories to put them into, what is the value of the Boolean function denoted by the map (\ref{MAP}) if the subspaces $\mathrm{ran}(\hat{Q})$ and $\mathrm{ran}(\hat{P})$ are neither ordered nor orthogonal? That is, what truth value should be assigned to the proposition $\mathcal{S}(\hat{Q})\mkern-1.5mu\lesseqgtr\mkern-1mu\mathcal{S}(\hat{P})$ in case the projection operators $\hat{Q}$ and $\hat{P}$ do not commute (are noncompatible)? In symbols,\smallskip

\begin{equation} \label{PR2} 
   {[\mkern-3.3mu[\mathcal{S}(\hat{Q})\mkern-1.5mu\lesseqgtr\mkern-1mu\mathcal{S}(\hat{P})]\mkern-3.3mu]}_b
   =
   b
   \left(
      {[\mkern-3.3mu[z]\mkern-3.3mu]}_v
      ,
      {[\mkern-3.3mu[w]\mkern-3.3mu]}_v
   \right)
   =
   \left\{
      \begingroup\SmallColSep
      \begin{array}{r l}
         1
         &
         \mkern3mu
         ,
         \mkern12mu
         {[\mkern-3.3mu[z]\mkern-3.3mu]}_v
         =
         1
         \\[5pt]
         0
         &
         \mkern3mu
         ,
         \mkern12mu
         {[\mkern-3.3mu[w]\mkern-3.3mu]}_v
         =
         1
         \\[5pt]
         ?
         &
         \mkern3mu
         ,
         \mkern12mu
         {[\mkern-3.3mu[z]\mkern-3.3mu]}_v
         =
         {[\mkern-3.3mu[w]\mkern-3.3mu]}_v
         =
         0
      \end{array}
      \endgroup   
   \right.
   \;\;\;\;  .
\end{equation}
\smallskip

\noindent To overcome the problem of an extra object, two simple workarounds are available.\\

\section{Birkhoff and von Neumann's workaround}  

\noindent A workaround proposed by Birkhoff and von Neumann \cite{Birkhoff, Piron} is to assume that the map (\ref{MAP}) is a non-injective surjective function. That is, it is not required that elements of $\mathbb{B}_2\times\mathbb{B}_2$ be unique; the Boolean function $b$ may associate two elements of $\mathbb{B}_2\times\mathbb{B}_2$ with one and the same element of $\mathbb{B}_2$. As a result,\smallskip

\begin{equation} \label{ABN} 
   b(0,0)
   \mkern5mu
   \text{is equal to either}
   \mkern5mu
   b(1,0)
   \mkern5mu
   \text{or}
   \mkern5mu
   b(0,1)
\;\;\;\;  .
\end{equation}
\smallskip

\noindent Suppose that $b(0,0)$ is equal to $b(0,1)$. Then, for example, if $x$ is true, $b$ returns 1, but in case $y$ is true or the conjunction ${\neg}x{\mkern2.5mu\sqcap\mkern2mu}{\neg}y$ is true, $b$ returns 0.\\

\noindent On the other hand, for any two mathematical statements, say $s_1$ and $s_2$, such that $s_1{\mkern2.5mu\sqcap\mkern2.5mu}s_2 = \bot$ even though $s_1{\mkern2.5mu\sqcup\mkern2.5mu}s_2 \neq \top$ (where $\bot$ and $\top$ denote arbitrary contradiction and tautology, respectively, i.e., propositions which are false or true in any nonzero state: ${[\mkern-3.3mu[\bot]\mkern-3.3mu]}_b \equiv 0$ and ${[\mkern-3.3mu[\top]\mkern-3.3mu]}_b \equiv 1$), the following logical biconditionals hold:\smallskip

\begin{equation}  
   s_{1,2}
   \mkern2.5mu
   \sqcup
   \mkern2.5mu
   \left(
      {\neg}s_1
      \mkern2.5mu
      \sqcap
      \mkern2.5mu
      {\neg}s_2
   \right)
   \mkern2.5mu
   \iff
   \mkern2.5mu
   {\neg}s_{2,1}
\;\;\;\;  .
\end{equation}
\smallskip

\noindent For this reason, the application of the assumption (\ref{ABN}) to (\ref{PR1}) makes the truth value of the experimental atomic proposition $P$ subject to the truth value of the statement $x\mkern-3.3mu:\mkern2mu|\Psi\rangle\mkern-2.5mu\in\mkern-2mu\mathrm{ran}(\hat{P})$ or the statement $y\mkern-3.3mu:\mkern2mu|\Psi\rangle\mkern-2.5mu\in\mkern-2mu\mathrm{ker}(\hat{P})$ alone, but not both. E.g.,\smallskip

\begin{equation} \label{QL} 
   {[\mkern-3.3mu[P]\mkern-3.3mu]}_b
   =
   b
   \left(
      {[\mkern-3.3mu[x]\mkern-3.3mu]}_v
   \right)
   =
   \left\{
      \begingroup\SmallColSep
      \begin{array}{r l}
         1
         &
         \mkern3mu
         ,
         \mkern12mu
         {[\mkern-3.3mu[x]\mkern-3.3mu]}_v
         =
         1
         \\[5pt]
         0
         &
         \mkern3mu
         ,
         \mkern12mu

         {[\mkern-3.3mu[x]\mkern-3.3mu]}_v
         =
         0
      \end{array}
      \endgroup   
   \right.
   \;\;\;\;  .
\end{equation}
\smallskip

\noindent Likewise, the application of the assumption (\ref{ABN}) to (\ref{PR2}) makes the truth value of the proposition $\mathcal{S}(\hat{Q})\mkern-1.5mu\lesseqgtr\mkern-1mu\mathcal{S}(\hat{P})$ subject to the truth value of either the statement $z$ declaring the order by set inclusion for the subspaces $\mathrm{ran}(\hat{Q})$ and $\mathrm{ran}(\hat{P})$, or the statement $w$ asserting their orthogonality. Thus,\smallskip

\begin{equation} \label{QL1} 
   {[\mkern-3.3mu[\mathcal{S}(\hat{Q})\mkern-1.5mu\lesseqgtr\mkern-1mu\mathcal{S}(\hat{P})]\mkern-3.3mu]}_b
   =
   b
   \left(
      {[\mkern-3.3mu[z]\mkern-3.3mu]}_v
   \right)
   =
   \left\{
      \begingroup\SmallColSep
      \begin{array}{r l}
         1
         &
         \mkern3mu
         ,
         \mkern12mu
         {[\mkern-3.3mu[z]\mkern-3.3mu]}_v
         =
         1
         \\[5pt]
         0
         &
         \mkern3mu
         ,
         \mkern12mu
         {[\mkern-3.3mu[z]\mkern-3.3mu]}_v
         =
         0
      \end{array}
      \endgroup   
   \right.
   \;\;\;\;  .
\end{equation}
\smallskip

\noindent Despite being seemingly elegant, Birkhoff and von Neumann’s workaround is marred by controversy. Specifically, not only this workaround brings in the failure of the distributive law of propositional logic, it also creates a paradoxical situation, in which qubits are discriminated against many-valued quantum systems.\\

\noindent According to the valuation (\ref{QL1}), every pair of elements of the set $L(\mathcal{H})$ are either ordered or not ordered by set inclusion. This means that $L(\mathcal{H})$ is a poset, i.e., a set with a partial order.\\

\noindent Providing each pair $\{\mathrm{ran}(\hat{Q}),\mathrm{ran}(\hat{P})\}$ of the poset $L(\mathcal{H})$ has a meet $\mathrm{ran}(\hat{Q})\wedge\mathrm{ran}(\hat{P})$ and a join $\mathrm{ran}(\hat{Q})\vee\mathrm{ran}(\hat{P})$ defined (in accordance with \cite{Halmos}) by\smallskip

\begin{equation} \label{MEET} 
   \mathrm{ran}(\hat{Q})
   \wedge
   \mathrm{ran}(\hat{P})
   =
   \mathrm{ran}(\hat{Q})
   \cap
   \mathrm{ran}(\hat{P})
\;\;\;\;  ,
\end{equation}
\\[-36pt]

\begin{equation} \label{JOIN} 
   \mathrm{ran}(\hat{Q})
   \vee
   \mathrm{ran}(\hat{P})
   =
   \left(
      \mathrm{ker}(\hat{Q})
      \cap
      \mathrm{ker}(\hat{P})
   \right)^{\perp}
\;\;\;\;  ,
\end{equation}
\smallskip

\noindent where $\cap$ denotes set-theoretic intersection, the poset $L(\mathcal{H})$ is a lattice. However, this lattice is not distributive.\\

\noindent To see this, consider atomic propositions $Q$, $P_1$ and $P_2$ pertaining to qubits. Suppose that $P_1{\mkern2.5mu\sqcap\mkern2.5mu}P_2 = \bot$ but $P_1{\mkern2.5mu\sqcup\mkern2.5mu}P_2 = \top$. The propositions $P_1$ and $P_2$ are represented by the subspaces $\mathrm{ran}(\hat{P}_1)$ and $\mathrm{ran}(\hat{P}_2)$ whose accordant projection operators $\hat{P}_1$ and $\hat{P}_2$ are compatible and orthogonal: $\hat{P}_{1}\hat{P}_{2}=\hat{P}_{2}\hat{P}_{1}=0$. Taking into consideration their orthogonality, i.e.,\smallskip

\begin{equation}  
   \mathrm{ran}(\hat{P}_{1,2})
   =
   \mathrm{ker}(\hat{P}_{2,1})
\;\;\;\;   ,
\end{equation}
\smallskip

\noindent and making use of (\ref{JOIN}), one gets:\smallskip

\begin{equation}  
   \mathrm{ran}(\hat{P}_1)
   \vee
   \mathrm{ran}(\hat{P}_2)
   =
   \left(
      \mathrm{ker}(\hat{P}_1)
      \cap
      \mathrm{ker}(\hat{P}_2)
   \right)^{\perp}
   =
   \left(
      \mathrm{ran}(\hat{P}_2)
      \cap
      \mathrm{ker}(\hat{P}_2)
   \right)^{\perp}
   =
   \{0\}^{\perp}
   =
   \mathcal{H}
\;\;\;\;  .
\end{equation}
\smallskip

\noindent Suppose that the projection operator $\hat{Q}$ corresponding to the proposition $Q$ does not commute with $\hat{P}_1$ and $\hat{P}_2$. As a result of this, the intersection\smallskip

\begin{equation}  
   \mathrm{ran}(\hat{Q})
   \cap
   \mathrm{ran}(\hat{P}_{1,2})
   =
   \left\{
      |\varphi\rangle
      \mkern-1mu
      \in
      \mkern-1mu
      \mathcal{H}
      \textnormal{:}
      \mkern10mu
      |\varphi\rangle
      \mkern-1mu
      \in
      \mkern-1mu
      \mathrm{ran}(\hat{Q})
      \mkern5mu
      \text{and}
      \mkern5mu
      |\varphi\rangle
      \mkern-1mu
      \in
      \mkern-1mu
      \mathrm{ran}(\hat{P}_{1,2})
   \right\}
\;\;\;\;   
\end{equation}
\smallskip

\noindent is the zero subspace $\{0\}$ containing only the vector $|\varphi\rangle=0$. Given that the join of two zero subspaces is the zero subspace again, one finally finds that distributivity of $\wedge$ over $\vee$ does not hold:\smallskip

\begin{equation}  
   \bigvee_{k=1}^{2}
   \mathrm{ran}(\hat{Q})
   \wedge
   \mathrm{ran}(\hat{P}_{k})
   =
   \{0\}
\;\;\;\;  ,
\end{equation}
\\[-30pt]

\begin{equation}  
   \mathrm{ran}(\hat{Q})
   \wedge
   \bigvee_{k=1}^{2}
   \mathrm{ran}(\hat{P}_{k})
   =
   \mathrm{ran}(\hat{Q})
   \cap
   \mathcal{H}
   =
   \mathrm{ran}(\hat{Q})
\;\;\;\;  .
\end{equation}
\smallskip

\noindent On condition that logical conjunctions and disjunctions of experimental propositions relating to a quantum system are respectively characterized by meets and joins of elements of the lattice $L(\mathcal{H})$ representing those propositions, nondistributiveness of $L(\mathcal{H})$ entails the breakdown of the distributive law of propositional logic.\\

\noindent Suppose that a quantum system is prepared in the state determined by the truth value ${[\mkern-3.3mu[s]\mkern-3.3mu]}_v$ of some mathematical statement $s$ relating to the system. Consider the probability that if it were to be tested, the experimental proposition $P$ about this system would have the truth value of true. Let this probability be denoted by the value of the probability function $p_{{[\mkern-3.3mu[s]\mkern-3.3mu]}_v}$ at $P$, i.e., $p_{{[\mkern-3.3mu[s]\mkern-3.3mu]}_v}\mkern-3.5mu(P)$.\\

\noindent As it can be shown (see, for example, \cite{Hajek, Williamson}), if the probability function $p_{{[\mkern-3.3mu[s]\mkern-3.3mu]}_v}$ is a function that uniquely associates experimental propositions with real numbers 

\begin{equation}  
   p_{{[\mkern-3.3mu[s]\mkern-3.3mu]}_v}
   \mkern-5mu
   :
   \mkern2mu
   \mathbb{P}
   \to
   \mathbb{R}
   \;\;\;\;  ,
\end{equation}
\smallskip

\noindent where $\mathbb{P}$ and $\mathbb{R}$ stand respectively for the sets of experimental propositions and real numbers, and therewith $p_{{[\mkern-3.3mu[s]\mkern-3.3mu]}_v}$ satisfies the following constraints\smallskip

\begin{equation}  
   \forall
   P
   \in
   \mathbb{P}
   \textnormal{:}
   \mkern15mu
   p_{{[\mkern-3.3mu[s]\mkern-3.3mu]}_v}\mkern-3.5mu(P)
   \ge
   0
\;\;\;\;  ,
\end{equation}
\\[-26pt]

\begin{equation}  
   p_{{[\mkern-3.3mu[s]\mkern-3.3mu]}_v}\mkern-3.5mu(P)
   =
   \left\{
      \begingroup\SmallColSep
      \begin{array}{r l}
         1
         &
         \mkern3mu
         ,
         \mkern12mu
         {[\mkern-3.3mu[P]\mkern-3.3mu]}_b
         =
         1
         \\[5pt]
         0
         &
         \mkern3mu
         ,
         \mkern12mu
         {[\mkern-3.3mu[P]\mkern-3.3mu]}_b
         =
         0
      \end{array}
      \endgroup   
   \right.
   \;\;\;\;  ,
\end{equation}
\\[-26pt]

\begin{equation}  
   \forall
   P
   ,
   Q
   \in
   \mathbb{P}
   ,
   \mkern5mu
   P \sqcap Q
   =
   \bot
   \mkern3mu
   \textnormal{:}
   \mkern15mu
   p_{{[\mkern-3.3mu[s]\mkern-3.3mu]}_v}\mkern-3.5mu(P \sqcup Q)
   =
   p_{{[\mkern-3.3mu[s]\mkern-3.3mu]}_v}\mkern-3.5mu(P)
   +
   p_{{[\mkern-3.3mu[s]\mkern-3.3mu]}_v}\mkern-3.5mu(Q)
\;\;\;\;  ,
\end{equation}
\smallskip

\noindent then\smallskip

\begin{equation}  
   \forall
   P
   \in
   \mathbb{P}
   \mkern3mu
   \textnormal{:}
   \mkern15mu
   p_{{[\mkern-3.3mu[s]\mkern-3.3mu]}_v}\mkern-3.5mu(P)
   \in
   [0,1]
\;\;\;\;  ,
\end{equation}
\\[-31pt]

\begin{equation}  
   \forall
   P
   ,
   Q
   \in
   \mathbb{P}
   ,
   \mkern5mu
   P
   \mkern-7mu
   \iff
   \mkern-7mu
   Q
   \mkern3mu
   \textnormal{:}
   \mkern15mu
   p_{{[\mkern-3.3mu[s]\mkern-3.3mu]}_v}\mkern-3.5mu(P)
   =
   p_{{[\mkern-3.3mu[s]\mkern-3.3mu]}_v}\mkern-3.5mu(Q)
\;\;\;\;  .
\end{equation}
\smallskip

\noindent Consider the statements concerning qubits, namely, $x_1\mkern-3.3mu:\mkern2mu|\Psi\rangle\mkern-2.5mu\in\mkern-2mu\mathrm{ran}(\hat{P}_1)$ and $x_2\mkern-3.3mu:\mkern2mu|\Psi\rangle\mkern-2.5mu\in\mkern-2mu\mathrm{ran}(\hat{P}_2)$, where $\mathrm{ran}(\hat{P}_2) = \mathrm{ker}(\hat{P}_1)$ and so $y_1{\mkern-2mu\iff\mkern-1mu}x_2$. Suppose that the qubit is prepared in the state given by ${[\mkern-3.3mu[x_1]\mkern-3.3mu]}_v=0$. In this case, ${[\mkern-3.3mu[P_1]\mkern-3.3mu]}_b=0$ in accordance with the valuation (\ref{QL}); hence, the probability of the proposition $P_1$ being verified must be 0, or symbolically,\smallskip

\begin{equation}  
   p_{{[\mkern-3.3mu[x_1]\mkern-3.3mu]}_v=\mkern2.5mu{0}}\mkern-1mu(P_1)
   =
   0
\;\;\;\;  .
\end{equation}
\smallskip

\noindent Then again, if the statement $x_1$ is false, the statement $x_2$ can be either true or false. However, within Birkhoff and von Neumann’s workaround $b(0,0)=b(0,1)$, explicitly,\smallskip

\begin{equation}  
   b
   \Big(
      {[\mkern-3.3mu[x_1]\mkern-3.3mu]}_v
      =
      0
      ,
      {[\mkern-3.3mu[x_2]\mkern-3.3mu]}_v
      =
      0
   \Big)
   =
   b
   \Big(
      {[\mkern-3.3mu[x_1]\mkern-3.3mu]}_v
      =
      0
      ,
      {[\mkern-3.3mu[x_2]\mkern-3.3mu]}_v
      =
      1
   \Big)
\;\;\;\;  ,
\end{equation}
\smallskip

\noindent i.e., the case where both $x_1$ and $x_2$ are false is indistinguishable from the case where $x_1$ is false but $x_2$ is true. As a result,\smallskip

\begin{equation}  
   p_{{[\mkern-3.3mu[x_1]\mkern-3.3mu]}_v=\mkern2.5mu{0}}\mkern-1mu(P_2)
   =
   p_{{[\mkern-3.3mu[x_2]\mkern-3.3mu]}_v=\mkern2.5mu{1}}\mkern-1mu(P_2)
   =
   1
\;\;\;\;  .
\end{equation}
\smallskip

\noindent Now, let the qubit be prepared in the state given by ${[\mkern-3.3mu[x_1]\mkern-3.3mu]}_v=1$. Then, ${[\mkern-3.3mu[P_1]\mkern-3.3mu]}_b=1$ and ${[\mkern-3.3mu[P_2]\mkern-3.3mu]}_b=0$ thanks to $x_1{\mkern2mu\sqcap\mkern2mu}x_2=\bot$; consequently,\smallskip

\begin{equation}  
   p_{{[\mkern-3.3mu[x_1]\mkern-3.3mu]}_v=\mkern2.5mu{1}}\mkern-1mu(P_1)
   =
   1
\;\;\;\;  ,
\end{equation}
\\[-33pt]

\begin{equation}  
   p_{{[\mkern-3.3mu[x_1]\mkern-3.3mu]}_v=\mkern2.5mu{1}}\mkern-1mu(P_2)
   =
   p_{{[\mkern-3.3mu[x_2]\mkern-3.3mu]}_v=\mkern2.5mu{0}}\mkern-1mu(P_2)
   =
   0
\;\;\;\;  .
\end{equation}
\smallskip

\noindent Hence, for qubits, the probability function $p_{{[\mkern-3.3mu[s]\mkern-3.3mu]}_v}$, where $s=x_1{\mkern2mu\sqcup\mkern2.5mu}x_2$, is dispersion-free, i.e., this function has only the numbers 0 and 1 as domain:\smallskip

\begin{equation}  
   p_{{[\mkern-3.3mu[s]\mkern-3.3mu]}_v\in\mkern2.5mu{\{0,1\}}}
   \mkern-3mu
   \left(
      P_{1,2}
   \mkern-1mu
   \right)
   \in
   \{0,1\}
\;\;\;\;  .
\end{equation}
\smallskip

\noindent By contrast, if a $n$-level quantum system, which is characterized by $n>2$ experimental atomic propositions $P_1$, $\dots$, $P_n$, such that $P_k{\mkern2mu\sqcap\mkern2mu}P_{l{\neq}k}=\bot$ and $\sqcup_{k=1}^{n}P_k=\top$, is prepared in the state given by ${[\mkern-3.3mu[x_1]\mkern-3.3mu]}_v=0$, one gets the sum\smallskip

\begin{equation}  
   p_{{[\mkern-3.3mu[x_1]\mkern-3.3mu]}_v=\mkern2.5mu{0}}
   \mkern-1mu
      \big(
         \mkern-2mu
         \bigsqcup_{k=1}^{n}P_k
         \mkern-0.5mu
      \big)
   =
   \sum_{k=2}^{n}
      p_{{[\mkern-3.3mu[x_1]\mkern-3.3mu]}_v=\mkern2.5mu{0}}\mkern-1mu(P_k)
   =
   1
\;\;\;\;  ,
\end{equation}
\smallskip

\noindent whose summands $p_{{[\mkern-3.3mu[x_1]\mkern-3.3mu]}_v=\mkern2.5mu{0}}\mkern-1mu(P_k)$ are not required be only 0 or 1. So, in general,\smallskip

\begin{equation}  
   p_{{[\mkern-3.3mu[s]\mkern-3.3mu]}_v\in\mkern2.5mu{\{0,1\}}}
   \mkern-3mu
   \left(
      P_{1,2,3,{\dots},n}
   \mkern-1mu
   \right)
   \in
   [0,1]
\;\;\;\;  ,
\end{equation}
\smallskip

\noindent where $s=x_k{\mkern2mu\sqcup\mkern2.5mu}y_k$, $k\in\{1,2,3,{\dots},n\}$. Accordingly, one can conclude that Birkhoff and von Neumann’s workaround discriminates against qubits.\\

\noindent In the literature, one can find an ample amount of works dedicated to erasing this discrimination. A common approach exhibited in those works is to come up with some assumption(s) – additional to (\ref{ABN}) – which would facilitate bringing qubits in the domain of a Gleason-type theorem presenting a mathematical motivation for continuous quantum probabilities. So far, this approach has yielded no satisfactory resolution of the qubit discrimination: Some of the added assumptions are disputed whereas the others lead to new problems (see the critical analysis of the selected papers in \cite{Zela, Hall}).\\

\noindent Anyhow, a continuous flow of publications on the extension of Gleason's theorem to the 2-dimensional case can be viewed as a sign that, most likely, there is no easy solution for the discrimination against qubits emerging from Birkhoff and von Neumann's workaround.\\

\section{The workaround based on the assumption of a partial Boolean function}  

\noindent The alternative workaround allowing to overcome the problem of an extra object is to assume that the map (\ref{MAP}) is a partial Boolean function. That is, the function $b$ is not forced to map every element of $\mathbb{B}_2\times\mathbb{B}_2$ to an element of $\mathbb{B}_2$ but only some 2-member subset of $\mathbb{B}_2\times\mathbb{B}_2$. Consequently, the pair $(0,0)$ has no association with the elements of $\mathbb{B}_2$ and therefore\smallskip

\begin{equation} \label{ASV} 
   b(0,0)
   \mkern5mu
   \text{is not defined}
\;\;\;\;  .
\end{equation}
\smallskip

\noindent According to this assumption (which can be called \emph{the assumption of a partial Boolean function}), the truth value of the experimental atomic proposition $P$ in the pure state described by the vector $|\Psi\rangle$ is determined by\smallskip

\begin{equation} \label{SV1} 
   {[\mkern-3.3mu[P]\mkern-3.3mu]}_b
   =
   b
   \left(
      {[\mkern-3.3mu[x]\mkern-3.3mu]}_v
      ,
      {[\mkern-3.3mu[y]\mkern-3.3mu]}_v
   \right)
   =
   \left\{
      \begingroup\SmallColSep
      \begin{array}{r l}
         1
         &
         \mkern3mu
         ,
         \mkern12mu
        {[\mkern-3.3mu[x]\mkern-3.3mu]}_v
         =
         1
         \\[5pt]
         0
         &
         \mkern3mu
         ,
         \mkern12mu
         {[\mkern-3.3mu[y]\mkern-3.3mu]}_v
         =
         1
         \\[5pt]
         0/0
         &
         \mkern3mu
         ,
         \mkern12mu
         {[\mkern-3.3mu[x]\mkern-3.3mu]}_v
         =
         {[\mkern-3.3mu[y]\mkern-3.3mu]}_v
         =
         0
      \end{array}
      \endgroup   
   \right.
   \;\;\;\;  ,
\end{equation}
\smallskip

\noindent where $0/0$ denotes a truth-value gap. This gap signifies that in the case where the vector $|\Psi\rangle$ is a superposition of vectors belonging to $\mathrm{ran}(\hat{P})$ and $\mathrm{ker}(\hat{P})$, the proposition $P$ has no truth value at all.\\

\noindent In the same way, the truth value of the proposition $\mathcal{S}(\hat{Q})\mkern-1.5mu\lesseqgtr\mkern-1mu\mathcal{S}(\hat{P})$ is determined by\smallskip

\begin{equation}  
   {[\mkern-3.3mu[\mathcal{S}(\hat{Q})\mkern-1.5mu\lesseqgtr\mkern-1mu\mathcal{S}(\hat{P})]\mkern-3.3mu]}_b
   =
   b
   \left(
      {[\mkern-3.3mu[z]\mkern-3.3mu]}_v
      ,
      {[\mkern-3.3mu[w]\mkern-3.3mu]}_v
   \right)
   =
   \left\{
      \begingroup\SmallColSep
      \begin{array}{r l}
         1
         &
         \mkern3mu
         ,
         \mkern12mu
         {[\mkern-3.3mu[z]\mkern-3.3mu]}_v
         =
         1
         \\[5pt]
         0
         &
         \mkern3mu
         ,
         \mkern12mu
         {[\mkern-3.3mu[w]\mkern-3.3mu]}_v
         =
         1
         \\[5pt]
         0/0
         &
         \mkern3mu
         ,
         \mkern12mu
         {[\mkern-3.3mu[z]\mkern-3.3mu]}_v
         =
         {[\mkern-3.3mu[w]\mkern-3.3mu]}_v
         =
         0
      \end{array}
      \endgroup   
   \right.
   \;\;\;\;  .
\end{equation}
\smallskip

\noindent Compliant with the above valuation, there are pairs $\{\mathrm{ran}(\hat{Q}),\mathrm{ran}(\hat{P})\}$ in $L(\mathcal{H})$ which are neither permitted nor prohibited to be ordered by set inclusion (or reverse set inclusion). This means that the set $L(\mathcal{H})$ is not a poset.\\

\noindent Still, for the subsets $\mathcal{S}(\hat{Q})$ and $\mathcal{S}(\hat{P})$ whose accordant projection operators $\hat{Q}$ and $\hat{P}$ are compatible, the proposition $\mathcal{S}(\hat{Q})\mkern-1.5mu\lesseqgtr\mkern-1mu\mathcal{S}(\hat{P})$ is either true or false. That is, either the subspace $\mathrm{ran}(\hat{Q})$ precedes (follows) the subspace $\mathrm{ran}(\hat{P})$ in the ordering by set inclusion, or the subspace $\mathrm{ran}(\hat{Q})$ is orthogonal to the subspace $\mathrm{ran}(\hat{P})$. As a result, the meet $\mathrm{ran}(\hat{Q})\wedge\mathrm{ran}(\hat{P})$ and the join $\mathrm{ran}(\hat{Q})\vee\mathrm{ran}(\hat{P})$ are defined in that case. E.g., giving the stipulations (\ref{MEET}) and (\ref{JOIN}), one gets\smallskip

\begin{equation} \label{M1} 
   \mathrm{ran}(\hat{Q})\wedge\mathrm{ran}(\hat{P})
   =
   \mathrm{ran}(\hat{Q})
   \mkern15mu
   \text{if}
   \mkern15mu
   \mathrm{ran}(\hat{Q})
   \subseteq
   \mathrm{ran}(\hat{P})
   \;\;\;\;  ,
\end{equation}
\\[-35pt]

\begin{equation} \label{M2} 
   \mathrm{ran}(\hat{Q})\wedge\mathrm{ran}(\hat{P})
   =
   \mathrm{ran}(\hat{P})
   \mkern15mu
   \text{if}
   \mkern15mu
   \mathrm{ran}(\hat{P})
   \subseteq
   \mathrm{ran}(\hat{Q})
   \;\;\;\;  ,
\end{equation}
\\[-35pt]

\begin{equation} \label{M0} 
   \mathrm{ran}(\hat{Q})\wedge\mathrm{ran}(\hat{P})
   =
   \{0\}
   \mkern15mu
   \text{if}
   \mkern15mu
   \mathrm{ran}(\hat{Q})
   \subseteq
   \mathrm{ker}(\hat{P})
   \mkern15mu
   \text{or}
   \mkern15mu
   \mathrm{ran}(\hat{P})
   \subseteq
   \mathrm{ker}(\hat{Q})
\;\;\;\;  .
\end{equation}
\smallskip

\noindent In opposition, if projection operators $\hat{Q}$ and $\hat{P}$ are noncompatible, the proposition $\mathcal{S}(\hat{Q})\mkern-1.5mu\lesseqgtr\mkern-1mu\mathcal{S}(\hat{P})$ has a truth-value gap. In that case, the subspaces $\mathrm{ran}(\hat{Q})$ and $\mathrm{ran}(\hat{P})$ neither precede each other in the ordering, nor they are orthogonal to each other. Consequently, binary operations (\ref{M1}), (\ref{M2}) and (\ref{M0}) are not defined on them. One can infer from this fact that the logical conjunction and disjunction of experimental propositions $Q$ and $P$ associated with noncompatible projection operators $\hat{Q}$ and $\hat{P}$ are not defined either.\\

\noindent Let us reexamine the atomic propositions $Q$, $P_1$ and $P_2$ relating to qubits. Consider the statements: $x_{1,2}\mkern-3.3mu:\mkern2mu|\Psi\rangle\mkern-2.5mu\in\mkern-2mu\mathrm{ran}(\hat{P}_{1,2})$ and $y_{1,2}\mkern-3.3mu:\mkern2mu|\Psi\rangle\mkern-2.5mu\in\mkern-2mu\mathrm{ker}(\hat{P}_{1,2})$. It is easy to see that $x_{1,2}{\mkern-2mu\iff\mkern-1mu}y_{2,1}$.\\

\noindent Suppose that the qubit is prepared in the state given by ${[\mkern-3.3mu[x_1]\mkern-3.3mu]}_v=1$. In agreement with (\ref{SV1}), the probability that in this case $P_{1,2}$ would be verified is\smallskip

\begin{equation}  
   p_{{[\mkern-3.3mu[x_1]\mkern-3.3mu]}_v=\mkern2.5mu{[\mkern-3.3mu[y_2]\mkern-3.3mu]}_v=\mkern2.5mu{1}}\mkern-1mu(P_1)
   =
   1
\;\;\;\;  ,
\end{equation}
\\[-31pt]

\begin{equation}  
   p_{{[\mkern-3.3mu[x_1]\mkern-3.3mu]}_v=\mkern2.5mu{[\mkern-3.3mu[y_2]\mkern-3.3mu]}_v=\mkern2.5mu{1}}\mkern-1mu(P_2)
   =
   0
\;\;\;\;  .
\end{equation}
\smallskip

\noindent Likewise, if the qubit is prepared in the state given by ${[\mkern-3.3mu[y_1]\mkern-3.3mu]}_v=1$, this probability becomes\smallskip

\begin{equation}  
   p_{{[\mkern-3.3mu[y_1]\mkern-3.3mu]}_v=\mkern2.5mu{[\mkern-3.3mu[x_2]\mkern-3.3mu]}_v=\mkern2.5mu{1}}\mkern-1mu(P_1)
   =
   0
\;\;\;\;  ,
\end{equation}
\\[-31pt]

\begin{equation}  
   p_{{[\mkern-3.3mu[y_1]\mkern-3.3mu]}_v=\mkern2.5mu{[\mkern-3.3mu[x_2]\mkern-3.3mu]}_v=\mkern2.5mu{1}}\mkern-1mu(P_2)
   =
   1
\;\;\;\;  .
\end{equation}
\smallskip

\noindent In case the qubit is prepared in a superposition of vectors belonging to the subspaces $\mathrm{ran}(\hat{P}_{1,2})$ and $\mathrm{ker}(\hat{P}_{1,2})$, its state is given by ${[\mkern-3.3mu[x_{1,2}]\mkern-3.3mu]}_v={[\mkern-3.3mu[y_{1,2}]\mkern-3.3mu]}_v=0$ and so the propositions $P_1$ and $P_2$ have no truth values. On condition that the truth values of false, 0, and true, 1, correspond to the endpoints of the unit interval $[0,1]$, the probability that either of these propositions would be verified in the said case can be neither 0 nor 1. This implies\smallskip

\begin{equation}  
   p_{{[\mkern-3.3mu[x_{1,2}]\mkern-3.3mu]}_v=\mkern2.5mu{[\mkern-3.3mu[y_{1,2}]\mkern-3.3mu]}_v=\mkern2.5mu{0}}
   \mkern-3mu
   \left(
      P_{1,2}
   \mkern-1mu
   \right)
   \in
   (0,1)
\;\;\;\;  .
\end{equation}
\smallskip

\noindent Providing $s=x_1{\mkern2mu\sqcup\mkern2.5mu}x_2$, one gets\smallskip

\begin{equation}  
   p_{{[\mkern-3.3mu[s]\mkern-3.3mu]}_v\mkern2.5mu\in\{0,1\}}
   \mkern-3mu
   \left(
      P_{1,2}
   \mkern-1mu
   \right)
   \in
   [0,1]
\;\;\;\;  .
\end{equation}
\smallskip

\noindent It can be shown in a similar manner that for any $n$-level quantum system characterized by $n>2$ experimental atomic propositions $P_1$, $P_2$, $P_3$, $\dots$ satisfying the conditions $P_k{\mkern2mu\sqcap\mkern2mu}P_{l{\neq}k}=\bot$ and $\sqcup_{k=1}^{n}P_k=\top$, the following holds\smallskip

\begin{equation}  
   p_{{[\mkern-3.3mu[s]\mkern-3.3mu]}_v\mkern2.5mu\in\{0,1\}}
   \mkern-3mu
   \left(
      P_{1,2,3,{\dots},n}
   \mkern-1mu
   \right)
   \in
   [0,1]
\;\;\;\;  ,
\end{equation}
\smallskip

\noindent where $s=x_k{\mkern2mu\sqcup\mkern2.5mu}y_k$, $k\in\{1,2,3,{\dots},n\}$. Therefore, the workaround based on the assumption of a partial Boolean function does not discriminates against qubits.\\

\noindent Quite the opposite, as the next section is to show, this workaround gives rise to a distinction in favor of qubits.\\

\section{Operational discrimination between qubits and many-level systems}  

\noindent Let us show first that the assignment of truth values to experimental atomic propositions pertaining to a quantum system in a pure quantum state can be treated as a computational problem, namely, the inspection of solvability of the corresponding systems of linear equations.\\

\noindent Consider the quantum system associated with the separable Hilbert space $\mathcal{H}$ of finite dimension $n\ge2$. Let the projection operator $\hat{P}$ acting on $\mathcal{H}$ be encoded by the complex ${n}\times{n}$ matrix $\mathbf{M}(\hat{P})$\smallskip

\begin{equation}  
   \mathbf{M}(\hat{P})
   =
   \mkern-2mu
   \left[
      \begingroup\SmallColSep
      \begin{array}{c c c}
         M_{11}
         &
         \cdots
         &
         M_{1n}
         \\
         \vdots
         &
         \ddots
         &
         \vdots
         \\
         M_{n1}
         &
         \cdots
         &
         M_{nn}
      \end{array}
      \endgroup
   \right]
   \in
   \mathbb{C}^{{n}\times{n}}
   \;\;\;\;   
\end{equation}
\smallskip

\noindent whose entries $M_{ij}$ are defined by the expression\smallskip

\begin{equation}  
   M_{ij}
   =
   \langle{e_i}|\hat{P}|{e_j}\rangle
   \;\;\;\;  ,
\end{equation}
\smallskip

\noindent where $|{e_j}\rangle$ are vectors of an arbitrary orthonormal basis $\{|{e_j}\rangle\}$, $\langle{e_i}|{e_j}\rangle=\delta_{ij}$, for $\mathcal{H}$. The span of the column vectors\smallskip

\begin{equation}  
   \mathbf{M}_j
   =
   \left(
      M_{ij}
   \right)_{i=1}^{n}
   \in
   \mathbb{C}^{{n}\times{1}}
   \;\;\;\;  ,
\end{equation}
\smallskip

\noindent is the range of the matrix $\mathbf{M}(\hat{P})$; explicitly,\smallskip

\begin{equation}  
   \mathrm{ran}(\mathbf{M}(\hat{P}))
   =
   \left\{
         c_1
         ,
         \dots
         ,
         c_n
         \in
         \mathbb{C}
         :
         \mkern10mu
         c_1
         \mathbf{M}_{1}
         +
         \dots
         +
         c_n
         \mathbf{M}_{n}
   \right\}
   \;\;\;\;  ,
\end{equation}
\smallskip

\noindent where either $(\mathbf{M}_{1},\dots,\mathbf{M}_{n})$ is a basis for $\mathrm{ran}(\mathbf{M}(\hat{P}))$ or some $\mathbf{M}_{j}$ can be removed to obtain a basis for $\mathrm{ran}(\mathbf{M}(\hat{P}))$.\\

\noindent From here, one can make the following observation: The truth of the mathematical statement $|\Psi\rangle\mkern-2.5mu\in\mkern-2mu\mathrm{ran}(\hat{P})$ means the solvability of the system of linear equations\smallskip

\begin{equation}  
   \mathbf{RX}
   =
   \mathbf{\Psi}
    \;\;\;\;  ,
\end{equation}
\smallskip

\noindent where $\mathbf{\Psi}$ is the column vector\smallskip

\begin{equation}  
   \mathbf{\Psi}
   =
   \left(
      \psi_i
   \right)_{i=1}^{n}
   \in
   \mathbb{C}^{{n}\times{1}}
   \;\;\;\;  ,
\end{equation}
\smallskip

\noindent which contains the components $\psi_{i}$ of the unit vector $|\Psi\rangle$ with respect to the chosen basis $\{|{e_j}\rangle\}$, that is,\smallskip

\begin{equation}  
   \psi_{i}
   =
   \langle{e_i}|\Psi\rangle
   \;\;\;\;  ,
\end{equation}
\smallskip

\noindent while $\mathbf{X}$ is the column vector with $m$ unknowns $x_1$, $\dots$, $x_m$, which are put in the place of weights $c_1$, $\dots$, $c_m$ for the linearly independent column vectors $\mathbf{M}_{1}$, $\dots$, $\mathbf{M}_{m}$, so that\smallskip

\begin{equation}  
   \mathbf{RX}
   \equiv
   \left[
      \begingroup\SmallColSep
      \begin{array}{c}
         M_{11}  \\
         \vdots  \\
         M_{n1} 
      \end{array}
      \endgroup
   \right]
   \mkern-4mu
   x_1
   +
   \cdots
   +
   \left[
      \begingroup\SmallColSep
      \begin{array}{c}
         M_{1m}  \\
         \vdots  \\
         M_{nm} 
      \end{array}
      \endgroup
   \right]
   \mkern-4mu
   x_m
    \;\;\;\;  .
\end{equation}
\smallskip

\noindent Along the same lines, the kernel of the matrix $\mathbf{M}(\hat{P})$ can be presented as the span of the column vectors $\mathbf{I}_j-\mathbf{M}_j$\smallskip

\begin{equation}  
   \mathrm{ker}(\mathbf{M}(\hat{P}))
   =
   \left\{
         c_1
         ,
         \dots
         ,
         c_n
      \mkern-2mu
      \in
      \mkern-2mu
      \mathbb{C}
      :
      \mkern10mu
      c_1
      \left(
         \mathbf{I}_{1}
         -
         \mathbf{M}_{1}
      \right)
      +
      \dots
      +
      c_n
      \left(
         \mathbf{I}_{n}
         -
         \mathbf{M}_{n}
      \right)
   \right\}      
   \;\;\;\;  ,
\end{equation}
\smallskip

\noindent where\smallskip

\begin{equation}  
   \mathbf{I}_{j}
   =
   \left(
      \delta_{ij}
   \right)_{i=1}^{n}
   \in
   \mathbb{C}^{{n}\times{1}}
   \;\;\;\;   
\end{equation}
\smallskip

\noindent and either $(\mathbf{I}_{1}-\mathbf{M}_{1},\dots,\mathbf{I}_{n}-\mathbf{M}_{n})$ is a basis for $\mathrm{ker}(\mathbf{M}(\hat{P}))$ or some $\mathbf{I}_{j}-\mathbf{M}_{j}$ can be removed to obtain a basis for $\mathrm{ker}(\mathbf{M}(\hat{P}))$. In this way, to decide whether $|\Psi\rangle$ belongs to $\mathrm{ker}(\hat{P})$ means to answer the question whether the following system of linear equations has at least one solution:\smallskip

\begin{equation}  
   \mathbf{KX}
   =
   \mathbf{\Psi}
    \;\;\;\;  .
\end{equation}
\smallskip

\noindent In the above, $\mathbf{X}$ is the column vector with $k$ unknowns $x_1$, $\dots$, $x_k$, which take the place of weights $c_1$, $\dots$, $c_k$ for the linearly independent column vectors $\mathbf{I}_{j}-\mathbf{M}_{j}$, so that\smallskip

\begin{equation}  
   \mathbf{KX}
   \equiv
   \left[
      \begingroup\SmallColSep
      \begin{array}{r}
         1-M_{11}  \\
         \vdots     \\
         -M_{n1} 
      \end{array}
      \endgroup
   \right]
   \mkern-4mu
   x_1
   +
   \cdots
   +
   \left[
      \begingroup\SmallColSep
      \begin{array}{r}
         \vdots                 \\
         \delta_{ij}-M_{ij} \\
         \vdots 
      \end{array}
      \endgroup
   \right]
   \mkern-4mu
   x_j
   +
   \cdots
   +
   \left[
      \begingroup\SmallColSep
      \begin{array}{r}
         -M_{1n}  \\
         \vdots   \\
         1-M_{nk} 
      \end{array}
      \endgroup
   \right]
   \mkern-4mu
   x_k
    \;\;\;\;  .
\end{equation}
\smallskip

\noindent Recall that the ket equation $\hat{P}|\Psi\rangle=|\Psi\rangle$ corresponds to the matrix equation\smallskip

\begin{equation}  
   \mathbf{M}(\hat{P})
   \mathbf{\Psi}
   =
   \mathbf{\Psi}
   \;\;\;\;  ,
\end{equation}
\smallskip

\noindent at the same time as the bra equation $\langle\Psi|=\langle\Psi|\hat{P}$ relates with the matrix equation\smallskip

\begin{equation}  
   \mathbf{\Psi}^{\dagger}
   =
   \mathbf{\Psi}^{\dagger}\mathbf{M}(\hat{P})
   \;\;\;\;  ,
\end{equation}
\smallskip

\noindent where $\mathbf{\Psi}^{\dagger}$ denotes the row vector $\mathbb{C}^{{1}\times{n}}$, whose $i^{\text{th}}$  entry is $\psi_i=\langle\Psi|e_i\rangle$. Therefore\smallskip

\begin{equation}  
   \mathbf{M}(\hat{P})
   =
   \mathbf{\Psi}
   \mathbf{\Psi}^{\dagger}
   \;\;\;\;  ,
\end{equation}
\smallskip

\noindent meaning that the rank of the matrix $\mathbf{M}(\hat{P})$ is 1 (see the proof, for example, in \cite{Mirsky}). Consistent with the rank-nullity theorem \cite{Friedberg}, this implies:\smallskip

\begin{equation}  
   \mathrm{Rank}(\mathbf{M}(\hat{P}))
   =
   \dim
   \mkern-2.5mu
   \left(
      \mathrm{ran}(\mathbf{M}(\hat{P}))
   \right)
   =
   m
   =
   1
   \;\;\;\;  ,
\end{equation}
\\[-32pt]

\begin{equation}  
   \mathrm{Nullity}(\mathbf{M}(\hat{P}))
   =
   \dim
   \mkern-2.5mu
   \left(
      \mathrm{ker}(\mathbf{M}(\hat{P}))
   \right)
   =
   k
   =
   n
   -
   1
   \;\;\;\;  .
\end{equation}
\smallskip

\noindent Let $U_{\mathbf{R}}$ and $U_{\mathbf{K}}$ be solution sets for the linear systems $\mathbf{RX}\mkern-3mu=\mkern-2mu\mathbf{\Psi}$ and $\mathbf{KX}\mkern-3mu=\mkern-2mu\mathbf{\Psi}$, specifically,\smallskip

\begin{equation}  
   U_{\mathbf{R}}
   =
   \left\{
      \mathbf{X}
      \mkern-2mu
      \in
      \mkern-2mu
      \mathbb{C}^1
      \mkern-3mu
      :
      \mkern7.5mu
      \mathbf{RX}
      =
      \mathbf{\Psi}
   \right\}
   \;\;\;\;  ,
\end{equation}
\\[-32pt]

\begin{equation}  
   U_{\mathbf{K}}
   =
   \left\{
      \mathbf{X}
      \mkern-2mu
      \in
      \mkern-2mu
      \mathbb{C}^{{(n-1)}\times{1}}
      \mkern-3mu
      :
      \mkern7.5mu
      \mathbf{KX}
      =
      \mathbf{\Psi}
   \right\}
   \;\;\;\;  .
\end{equation}
\smallskip

\noindent Then, the following equivalences of the mathematical statements hold:\smallskip

\begin{equation}  
   |\Psi\rangle
   \mkern-2.5mu
   \in
   \mkern-2mu
   \mathrm{ran}(\hat{P})
   \iff
   U_{\mathbf{R}}
   \mkern-3.5mu
   \neq
   \mkern-3.5mu
   \varnothing
   \;\;\;\;  ,
\end{equation}
\\[-32pt]

\begin{equation}  
   |\Psi\rangle
   \mkern-2.5mu
   \in
   \mkern-2mu
   \mathrm{ker}(\hat{P})
   \iff
   U_{\mathbf{K}}
   \mkern-3.5mu
   \neq
   \mkern-3.5mu
   \varnothing
   \;\;\;\;  .
\end{equation}
\smallskip

\noindent As a result, the statements $x$ and $y$ – whose truth values determine the truth value of the experimental atomic proposition $P$ in the formula (\ref{SV1}) – can be set forth as follows:\smallskip

\begin{equation}  
   x
   \mkern-3mu
   :
   \mkern3.5mu
   U_{\mathbf{R}}
   \mkern-3.5mu
   \neq
   \mkern-3.5mu
   \varnothing
   \;\;\;\;  ,
\end{equation}
\\[-32pt]

\begin{equation}  
   y
   \mkern-3mu
   :
   \mkern3.5mu
   U_{\mathbf{K}}
   \mkern-3.5mu
   \neq
   \mkern-3.5mu
   \varnothing
   \;\;\;\;  .
\end{equation}
\smallskip

\noindent The linear system $\mathbf{RX}\mkern-3mu=\mkern-2mu\mathbf{\Psi}$ has only one unknown, whereas the linear system $\mathbf{KX}\mkern-3mu=\mkern-2mu\mathbf{\Psi}$ has $n-1$ unknowns; so, if $n>2$, to verify $y$ (and thus to assign 0 to the proposition $P$) will likely take more operations than to verify $x$ (and, as a result, to assign 1 to $P$).\\

\noindent To confirm this intuition, let us introduce the work of a computation, $W$, that is, the total number of primitive operations performed in order to solve a computational problem at hand \cite{Cormen}.\\

\noindent Let  $[\cdot]_s$ stand for a metric used to gauge the performance of the computation concerning the mathematical statement $s$. For example, $[W]_x$ denote the work of the computation invested in verifying the statement $x$.\\

\noindent Let us estimate $[W]_x$. The linear system $\mathbf{RX}\mkern-3mu=\mkern-2mu\mathbf{\Psi}$ will be consistent and, hence, the set $U_{\mathbf{R}}$ will contain at least one solution, if the following condition holds:\smallskip

\begin{equation}  
   \forall
   j
   \in
   \left\{
      2,
      \dots
      ,
      n
   \right\}
   \mkern-2mu
   :
   \mkern8mu
   \psi_1
   M_{j.1}
   =
   \psi_j
   M_{11}
   \;\;\;\;    .
\end{equation}
\smallskip

\noindent Subsequently, to prove that ${[\mkern-3.3mu[x]\mkern-3.3mu]}_v=1$ requires $2(n-1)$ multiplications and $n-1$ comparisons (note that comparisons as well as elementary arithmetic operations can be regarded as primitive operations). Hence, one may write down\smallskip

\begin{equation}  
   [W]_x
   =
   O(n)
   \;\;\;\;    .
\end{equation}
\smallskip

\noindent To estimate $[W]_y$, one may present the linear system $\mathbf{KX}\mkern-3mu=\mkern-2mu\mathbf{\Psi}$ as the augmented matrix\smallskip

\begin{equation}  
   \left[
   \mathbf{K}
   \mkern-2mu
   \left|
   \mathbf{\Psi}
   \right.
   \right]
   =
   \left[
      \mkern-8mu
      \begingroup
      \begin{array}{c c c c | c}
         a_{11}
         &
         a_{12}
         &
         \cdots
         &
         a_{1.n-1}
         &
         a_{1.n}
         \\ 
         a_{21}
         &
         a_{22}
         &
         \cdots
         &
         a_{2.n-1}
         &
         a_{2.n}
         \\ 
         \vdots
         &
         \vdots
         &
         \ddots
         &
         \vdots
         &
         \vdots
         \\ 
         a_{n.1}
         &
         a_{n.2}
         &
         \cdots
         &
         a_{n.n-1}
         &
         a_{n.n}
     \end{array}
      \mkern-5mu
      \endgroup
   \right]
   \;\;\;\;  ,
\end{equation}
\smallskip

\noindent whose elements are\smallskip

\begin{equation} 
   a_{jl}
   =
   \left\{
      \begingroup\SmallColSep
      \begin{array}{r l}
         \delta_{jl}
         -
         M_{jl}
         &
         \mkern3mu
         ,
         \mkern12mu
         l
         <
         n
         \\[5pt]
         \psi_j
         &
         \mkern3mu
         ,
         \mkern12mu
         l
         =
         n
      \end{array}
      \endgroup
   \right.
   \;\;\;\;  .
\end{equation}
\smallskip

\noindent Then, one may use a Gaussian-type elimination algorithm which is described below.\\

\noindent At the $i^{\text{th}}$  iteration of the algorithm, on condition that $a_{i.i}^{(i-1)}\neq0$, the augmented matrix $[\mathbf{K}\mkern-2mu\left|\mathbf{\Psi}\right.]^{(i)}$ is computed by the formula\smallskip

\begin{equation} \label{ALG} 
   \forall
   i
   \in
   \mathbb{N}
   ,
   \mkern2mu
   i
   \le
   (n-1)
   \mkern-2mu
   :
   \mkern12mu
   \left[\mathbf{K}\mkern-2mu\left|\mathbf{\Psi}\right.\right]^{(i)}
   =
   \left[\mathbf{K}\mkern-2mu\left|\mathbf{\Psi}\right.\right]^{(i-1)}
   \mkern-2mu
   -
   \mkern-2mu
   \left(
      \mkern-2mu
      \mathbf{A}_{i}^{(i-1)}
      \mkern-4mu
      \div
      a_{i.i}^{i-1}
   \right)
   \mkern-2mu
   \cdot
   \mkern-2mu
   \mathbf{B}_{i}^{(i-1)}
   \;\;\;\;    ,
\end{equation}
\smallskip

\noindent where\smallskip

\begin{equation}  
   \mathbf{A}_{i}^{(i-1)}
   =
   \left[\mathbf{K}\mkern-2mu\left|\mathbf{\Psi}\right.\right]^{(i-1)}
   \cdot
   \left(\delta_{ij}^{\mkern1.5mu{i}}\right)_{\substack{j=1\\l=1}}^{n}
   \;\;\;\;   ,
\end{equation}
\\[-22pt]

\begin{equation}  
   a_{i.{\mkern1.5mu}i}^{(i-1)}
   =
   \mathrm{tr}
   \mkern-2mu
   \left(
      \left[\mathbf{K}\mkern-2mu\left|\mathbf{\Psi}\right.\right]^{(i-1)}
      \cdot
      \left(\delta_{ij}^{\mkern1.5mu{i}}\right)_{\substack{j=1\\l=1}}^{n}
   \right)
   \;\;\;\;   ,
\end{equation}
\\[-22pt]

\begin{equation}  
   \mathbf{B}_{i}^{(i-1)}
   =
   \left(\delta_{ij}^{\mkern1.5mu{i}}\right)_{\substack{j=1\\l=1}}^{n}
   \cdot
   \left[\mathbf{K}\mkern-2mu\left|\mathbf{\Psi}\right.\right]^{(i-1)}
   \;\;\;\;   ,
\end{equation}
\\[-22pt]

\begin{equation} 
   \delta_{ij}^{\mkern1.5mu{i}}
   =
   \left\{
      \begingroup\SmallColSep
      \begin{array}{r l}
         1
         &
         \mkern3mu
         ,
         \mkern12mu
         i
         =
         j
         =
         l
         \\[5pt]
         0
         &
         \mkern3mu
         ,
         \mkern12mu
         \text{else}
      \end{array}
      \endgroup
   \right.
   \;\;\;\;  ,
\end{equation}
\smallskip

\noindent and it is postulated that $a_{jl}^{(0)}=a_{jl}$.\\

\noindent For $i=1$, one gets\smallskip

\begin{equation}  
   \left(\delta_{ij}^{\mkern1.5mu{1}}\right)_{\substack{j=1\\l=1}}^{n}
   =
   \left[
      \begingroup\SmallColSep
      \begin{array}{c c c c}
         1
         &
         0
         &
         \cdots
         &
         0
         \\ 
         0
         &
         0
         &
         \cdots
         &
         0
         \\ 
         \vdots
         &
         \vdots
         &
         \ddots
         &
         \vdots
         \\ 
         0
         &
         0
         &
         \cdots
         &
         0
     \end{array}
      \endgroup
   \right]
   \in
   \mathbb{C}^{n{\times}n}
   \;\;\;\;    ,
\end{equation}
\smallskip

\noindent so, after the first iteration of the algorithm, the augmented matrix of the linear system $\mathbf{KX}\mkern-3mu=\mkern-2mu\mathbf{\Psi}$ takes the form\smallskip

\begin{equation}  
   \left[
   \mathbf{K}
   \mkern-2mu
   \left|
   \mathbf{\Psi}
   \right.
   \right]^{(1)}
   =
   \left[
      \mkern-8mu
      \begingroup
      \begin{array}{c c c c | c}
         0
         &
         0
         &
         \cdots
         &
         0
         &
         0
         \\[3pt]   
         0
         &
         a_{22}^{(1)}
         &
         \cdots
         &
         a_{2.n-1}^{(1)}
         &
         a_{2.n}^{(1)}
         \\[3pt]   
         \vdots
         &
         \vdots
         &
         \ddots
         &
         \vdots
         &
         \vdots
         \\[3pt]   
         0
         &
         a_{n.2}^{(1)}
         &
         \cdots
         &
         a_{n.n-1}^{(1)}
         &
         a_{n.n}^{(1)}
     \end{array}
      \mkern-5mu
      \endgroup
   \right]
   \;\;\;\;    .
\end{equation}
\smallskip

\noindent If $i=n-1$, then\smallskip

\begin{equation}  
   \left(\delta_{ij}^{\mkern1.5mu{n-1}}\right)_{\substack{j=1\\l=1}}^{n}
   =
   \left[
      \begingroup\SmallColSep
      \begin{array}{c c c c c}
         0
         &
          \cdots
         &
         0
         &
         0
         &
         0
         \\ 
         \vdots
         &
         \ddots
         &
         \vdots
         &
         \vdots
         &
         \vdots
         \\ 
         0
         &
          \cdots
         &
         0
         &
         1
         &
         0
         \\ 
         0
         &
         \cdots
         &
         0
         &
         0
         &
         0
     \end{array}
      \endgroup
   \right]
   \in
   \mathbb{C}^{n{\times}n}
   \;\;\;\;    ,
\end{equation}
\smallskip

\noindent and so, after the final iteration, the augmented matrix becomes\smallskip

\begin{equation}  
   \left[
   \mathbf{K}
   \mkern-2mu
   \left|
   \mathbf{\Psi}
   \right.
   \right]^{(n-1)}
   =
   \left[
      \mkern-8mu
      \begingroup
      \begin{array}{c c c c | c}
         0
         &
         0
         &
         \cdots
         &
         0
         &
         0
         \\[3pt]   
         0
         &
         0
         &
         \cdots
         &
         0
         &
         0         \\[3pt]   
         \vdots
         &
         \vdots
         &
         \ddots
         &
         \vdots
         &
         \vdots
         \\[3pt]   
         0
         &
         0
         &
         \cdots
         &
         0
         &
         a_{n.n}^{(n-1)}
     \end{array}
      \endgroup
      \mkern-9mu
   \right]
   \;\;\;\;    .
\end{equation}
\smallskip

\noindent At this point, the following condition should be verified:\smallskip

\begin{equation}  
   a_{n.n}^{(n-1)}
   =
   0
   \;\;\;\;    .
\end{equation}
\smallskip

\noindent If it holds, the linear system $\mathbf{KX}\mkern-3mu=\mkern-2mu\mathbf{\Psi}$ is consistent and the statement $y\mkern-5mu:{\mkern2mu}U_{\mathbf{K}}\mkern-3.5mu\neq\mkern-3.5mu\varnothing$ is true; otherwise, ${[\mkern-3.3mu[y]\mkern-3.3mu]}_v=0$.\\

\noindent According to the formula (\ref{ALG}), at each iteration of the algorithm, the work of the computation of the elements $a_{jl}^{(i)}$ involves $O(n^2)$ elementary arithmetic operations. Consequently,\smallskip

\begin{equation}  
   [W]_y
   =
   O(n^3)
   \;\;\;\;    .
\end{equation}
\smallskip

\noindent To conclude so far, for experimental propositions pertaining to a many-level quantum system, the work of the computation spent in justifying their falsity is much greater than that spent in proving their truth. However, for a qubit, the work of the computation performed in order to validate either truth value of experimental propositions is one and the same. In symbols,\smallskip

\begin{equation}  
   n
   \gg
   2
   \mkern-2mu
   :
   \mkern8mu
   [W]_y
   \gg
   [W]_x
   \;\;\;\;    ,
\end{equation}
\\[-32pt]

\begin{equation}  
   n
   =
   2
   \mkern-2mu
   :
   \mkern8mu
   [W]_y
   =
   [W]_x
   \;\;\;\;    .
\end{equation}
\smallskip

\noindent In this sense, the workaround based on the assumption of a partial Boolean function discriminates in favor of qubits.\\

\section{Quantum parallel computing}  

\noindent Recall that the physical resource of a computation is the time needed for running it \cite{Homer}. To evaluate this resource, one can use the cost of the computation, $C$, i.e., the amount of time taken to solve the given computational problem \cite{Cormen}.\\

\noindent The discrimination in favor of qubits would be of no physical importance if the following relation were to be true:\smallskip

\begin{equation} \label{REL} 
   n
   \ge
   2
   \mkern-2mu
   :
   \mkern8mu
   \left[\mkern2mu{C}\mkern2mu\right]_y
   =
   \left[\mkern2mu{C}\mkern2mu\right]_x
   \;\;\;\;    ,
\end{equation}
\smallskip

\noindent where the sign ``='' means to express that the computational costs $\left[\mkern2mu{C}\mkern2mu\right]_y$ and $\left[\mkern2mu{C}\mkern2mu\right]_x$ – as functions of the dimension $n$ of the Hilbert space $\mathcal{H}$ – must have equal growth rates.\\

\noindent As to the cost $\left[\mkern2mu{C}\mkern2mu\right]_x$, one can set it down as follows\smallskip

\begin{equation}  
   \left[\mkern2mu{C}\mkern2mu\right]_x
   =
   \left[\mkern2mu{W}\mkern2mu\right]_x
   =
   O(n)
   \;\;\;\;    .
\end{equation}
\smallskip

\noindent On the other hand, it is reasonable to assume that the Gaussian elimination is an asymptotically optimal method for row reduction and, therefore, no algorithm verifying the statement $U_{\mathbf{K}}\mkern-3.5mu\neq\mkern-3.5mu\varnothing$ can have the number of primitive operations growing slower than $[W]_y=O(n^3)$ \cite{Robert}.\\

\noindent Hence, the relation (\ref{REL}) might be possible on condition that there is a model of computation which can permit execution of many primitive operations at once, so that the amount of time taken to solve the computational problem can be much less than the number of primitive operations needed to solve it, specifically,\smallskip

\begin{equation}  
   n
   \gg
   2
   \mkern-2mu
   :
   \mkern8mu
   \left[\mkern2mu{C}\mkern2mu\right]_y
   \ll
   \left[\mkern2mu{W}\mkern2mu\right]_y
   \;\;\;\;    .
\end{equation}
\smallskip

\noindent According to the postulates of quantum mechanics, such a model – which can be called a QPRAM (\emph{quantum parallel random-access machine}) – is plausible.\\

\noindent To prove this, consider a classical parallel random-access machine (PRAM) at first. It solves the computational problem at hand in time $T_p$ with $p$ classical processors such that the cost of parallel computation $C_p$ is defined by the product\smallskip

\begin{equation}  
   C_p
   =
   p
   \cdot
   T_p
   \;\;\;\;    ,
\end{equation}
\smallskip

\noindent where the running time $T_p$ is bounded from below by the work $W$, in other words, the amount of time used to run the computation sequentially (i.e., on a single processor):\smallskip

\begin{equation}  
   T_p
   \ge
   \frac{W}{p}
   \;\;\;\;     .
\end{equation}
\smallskip

\noindent From the law of work of classical parallel computing \cite{Casanova}, namely,\smallskip

\begin{equation}  
   C_p
   \ge
   W
   \;\;\;\;     ,
\end{equation}
\smallskip

\noindent it follows that the efficiency of a PRAM, denoted $E_p$, cannot exceed 1. Explicitly,\smallskip

\begin{equation}  
   E_p
   =
   \frac{W}{C_p}
   \le
   1
   \;\;\;\;    .
\end{equation}
\smallskip

\noindent In an analogous line of reasoning, a QPRAM solves the given computational problem in time $X_q$ on $q$ quantum processors; subsequently, $C_q$, the cost of quantum parallel computation, can be defined as the product\smallskip

\begin{equation}  
   C_q
   =
   q
   \cdot
   X_q
   \;\;\;\;    .
\end{equation}
\smallskip

\noindent But, unlike a classical parallel computer, a QPRAM can invoke ``quantum parallelism''. This refers to a capability to carry out many computations in a massive parallel way, i.e., at once \cite{Timpson, Horsman, Lini, Duwell}.\\

\noindent Let $|\Psi_{\text{PRAM}}^{\mkern3mu(i)}\rangle$ be the unit vector characterizing a PRAM with $p=n^2$ classical processors computing the elements $a_{jl}^{(i)}$ of the augmented matrix $[\mathbf{K}\mkern-2mu\left|\mathbf{\Psi}\right.]^{(i)}$ in accordance with the iteration formula (\ref{ALG}). The said vector describes the product state of the classical processors and therefore can be written as\smallskip

\begin{equation}  
   |\Psi_{\text{PRAM}}^{\mkern3mu(i)}\rangle
   =
   \bigotimes_{k=1}^{n^2}
   |{\mkern1.5mu}a_{k}^{(i)}\rangle
   \;\;\;\;    ,
\end{equation}
\smallskip

\noindent where each quantum state $|{\mkern1.5mu}a_{k}^{(i)}\rangle$ is assumed to correspond to the outcome of the computation of the element $a_{jl}^{(i)}$ on the $k^{\text{th}}$ classical processor in a way that\smallskip

\begin{equation}  
   \langle{a_m^{(i)}}|{\mkern1.5mu}{a_k^{(i)}}\rangle
   =
   \delta_{km}
   \;\;\;\;    ,
\end{equation}
\\[-32pt]

\begin{equation}  
   k
   =
   n\cdot(j-1)+l
   \;\;\;\;    ,
\end{equation}
\\[-32pt]

\begin{equation}  
   a_k^{(i)}
   =
   a_{jl}^{(i)}
   \;\;\;\;    .
\end{equation}
\smallskip

\noindent Providing each state $|{\mkern1.5mu}a_k^{(i)}\rangle$ is generated in one go, the vector $|\Psi_{\text{PRAM}}^{\mkern3mu(i)}\rangle$ can be computed in $T_p=O(1)$ time at the cost $C_p=O(n^2)$.\\

\noindent Recall that any irreversible computation can be presented as an evaluation of an invertible function \cite{Hagar}. In view of that, let’s assume that the iterated function on the elements of the augmented matrix $[\mathbf{K}\mkern-2mu\left|\mathbf{\Psi}\right.]^{(i)}$, namely,\smallskip

\begin{equation}  
   f
   \mkern-2mu
   :
   \mkern8mu
   [\mathbf{K}\mkern-2mu\left|\mathbf{\Psi}\right.]^{(i-1)}
   \longrightarrow
   [\mathbf{K}\mkern-2mu\left|\mathbf{\Psi}\right.]^{(i)}
   \;\;\;\;    ,
\end{equation}
\smallskip

\noindent can be presented as the state-generated oracle (or its inverse) that performs the unitary transformation $U$\smallskip

\begin{equation}  
   U
   \mkern-2mu
   :
   \mkern8mu
   |\Psi_{\text{PRAM}}^{\mkern3mu(i-1)}\rangle
   \longrightarrow
   |\Psi_{\text{PRAM}}^{\mkern3mu(i)}\rangle
   \;\;\;\;     
\end{equation}
\smallskip

\noindent (the term ``oracle'' indicates that the time taken by the unitary transformation $U$ is not included in the cost of the computation of the state $|\Psi_{\text{PRAM}}^{\mkern3mu(i)}\rangle$).\\

\noindent To convert the initial state of the PRAM, $|\Psi_{\text{PRAM}}^{\mkern3mu(0)}\rangle$, into its final state, $|\Psi_{\text{PRAM}}^{\mkern3mu(n-1)}\rangle$, requires $O(n)$ queries of the above oracle; therefore, the cost of the classical parallel computation necessary to the inspection of the solvability of the linear system $\mathbf{KX}\mkern-3mu=\mkern-2mu\mathbf{\Psi}$ is supposed to be cubic, i.e.,\smallskip

\begin{equation}  
   \left[\mkern2mu{C_p}\mkern2mu\right]_y
   =
   O(n^3)
   \;\;\;\;     .
\end{equation}
\smallskip

\noindent Different from the classical PRAM, its quantum counterpart can exist in a superposition of the states $|{\mkern1.5mu}a_k^{(i)}\rangle$. This means that the state of a $q$-processor quantum parallel computer checking whether the linear system $\mathbf{KX}\mkern-3mu=\mkern-2mu\mathbf{\Psi}$ has at least one solution can be described by the tensor product\smallskip

\begin{equation}  
   |\Psi_{\text{QPRAM}}^{\mkern3mu(i)}\rangle
   =
   \bigotimes_{m=1}^{q}
   |\Phi_{m}^{(i)}\rangle
   \;\;\;\;    ,
\end{equation}
\smallskip

\noindent in which $q$ does not depend on $n$ but the pure state of the $m^{\text{th}}$ quantum processor $|\Phi_{m}^{(i)}\rangle$ is the linear combination of $n^2$ states $|\mkern1.5mu{a_k^{(i)}}\rangle$, i.e.,\smallskip

\begin{equation}  
   |\Phi_{m}^{(i)}\rangle
   =
   \sum_{k=1}^{n^2}
   \alpha_{mk}|\mkern1.5mu{a_k^{(i)}}\rangle
   \;\;\;\;    ,
\end{equation}
\smallskip

\noindent where $\alpha_{mk}$ are superposition coefficients, i.e., complex numbers describing how much goes into each computation.\\

\noindent Since a superposition of the states $|\mkern1.5mu{a_k^{(i)}}\rangle$ is generated in only one go, every state $|\Phi_{m}^{(i)}\rangle$ can be computed in $X_q=O(1)$ time. Inasmuch as $q=O(1)$, this entails a constant cost of the computation, i.e., $q{\cdot}X_q=O(1)$. Then again, the initial state of the QPRAM can be converted into its final state by assessing the state-generated oracle $O(n)$ times, that is,\smallskip

\begin{equation}  
   |\Psi_{\text{QPRAM}}^{\mkern3mu(0)}\rangle
   \stackrel{U}{\longrightarrow}
   |\Psi_{\text{QPRAM}}^{\mkern3mu(1)}\rangle
   \stackrel{U}{\longrightarrow}
   \mkern9mu
   \dots
   \mkern3mu
   \stackrel{U}{\longrightarrow}
   |\Psi_{\text{QPRAM}}^{\mkern3mu(n-1)}\rangle
   \;\;\;\;     , 
\end{equation}
\smallskip

\noindent thus, the cost of the computation of $|\Psi_{\text{QPRAM}}^{\mkern3mu(n-1)}\rangle$ ends up linear, i.e., $O(n)$.

\noindent Let $|\mkern1.5mu{0}\rangle$ correspond to the zero outcome of the computation, that is, $a_{jl}^{(i)}=0$, for any $j$ and $l$. Furthermore, let all the superposition coefficients $\alpha_{mk}$ be alike. Then, if the solution set for the linear system $\mathbf{KX}\mkern-3mu=\mkern-2mu\mathbf{\Psi}$ is not empty, the final state of the $m^{\text{th}}$ quantum processor becomes\smallskip

\begin{equation}  
   |\Phi_{m}^{(n-1)}\rangle
   \propto
   |\mkern1.5mu{0}\rangle
   \;\;\;\;     , 
\end{equation}
\smallskip

\noindent otherwise, this state takes the form of the superposition\smallskip

\begin{equation}  
   |\Phi_{m}^{(n-1)}\rangle
   \propto
   (n^2-1)|\mkern1.5mu{0}\rangle
   +
   |\mkern1.5mu{a_{n.n}^{(n-1)}}\rangle
   \;\;\;\;     . 
\end{equation}
\smallskip

\noindent Suppose that after the final iteration, each quantum processor of the QPRAM shares the information with the corresponding counter $C_m$ whose initial state is described by\smallskip

\begin{equation}  
   |\mkern1.5mu{C_m}\rangle
   \propto
   |\mkern1.5mu{c_1}\rangle
   +
   (n^2-1)|\mkern1.5mu{c_2}\rangle
   \;\;\;\;     , 
\end{equation}
\smallskip

\noindent where $\langle{c_j}|\mkern3mu{c_l}\rangle=\delta_{jl}$. Also assume that there is a unitary transformation, say $U_C$, acting on the product state $|\Phi_{m}^{(n-1)}\rangle\otimes|\mkern3mu{C_m}\rangle$ of the composite system ``processor-counter'', such that $U_C$ correlates the state $|\mkern1.5mu{0}\rangle$ with the state $|\mkern1.5mu{c_1}\rangle$ as well as the state $|\mkern1.5mu{a_{n.n}^{(n-1)}}\rangle$ with the state $|\mkern1.5mu{c_2}\rangle$. This unitary operator (in other words, quantum gate) can be written as\smallskip

\begin{equation}  
   U_C(\tau)
   =
   \exp\mkern-3mu\left(-\frac{i\tau}{\hbar}H\right)
   \;\;\;\;     , 
\end{equation}
\smallskip

\noindent where $\tau$ is the time interval of the interaction between the quantum processor and the counter, while $H$ stands for the total Hamiltonian of the system ``processor-counter'', which can be effectively (i.e., during the interval of the interaction) presented by\smallskip

\begin{equation} 
   H
   =
   |\mkern1.5mu{0}\rangle
   \langle{0}\mkern1.5mu|
   \mkern-1.5mu\otimes\mkern-2mu
   |\mkern1.5mu{c_1}\rangle
   \langle\mkern1.5mu{c_1}|
   +
   |\mkern1.5mu{a_{n.n}^{(n-1)}}\rangle
   \langle\mkern1.5mu{a_{n.n}^{(n-1)}}|
   \mkern-1.5mu\otimes\mkern-2mu
   |\mkern1.5mu{c_2}\rangle
   \langle\mkern3mu{c_2}|
   \;\;\;\;    .
\end{equation}
\smallskip

\noindent Note that $\langle{0}|\mkern1.5mu{c_2}\rangle=\langle{c_2}|\mkern1.5mu{0}\rangle=0$ which entails $H=|\mkern1.5mu{0}\rangle\langle{0}|\mkern-1.5mu\otimes\mkern-2mu|\mkern1.5mu{c_1}\rangle\langle{c_1}\mkern-1.5mu|$ in case $a_{n.n}^{(n-1)}\mkern-5mu=0$.\\

\noindent The gate $U_C$ transforms the product state $|\Phi_{m}^{(n-1)}\rangle\otimes|\mkern3mu{C_m}\rangle$ into an entangled state as follows:\smallskip

\begin{equation} 
   U_C
   \mkern-2mu
   :
   \mkern8mu
   |\Phi_{m}^{(n-1)}\rangle
   \otimes
   |\mkern3mu{C_m}\rangle
   \longrightarrow
   |\Phi_{m}^{(n-1)}\mkern3mu{C_m}\rangle
   \;\;\;\;    ,
\end{equation}
\smallskip

\noindent where\smallskip

\begin{equation} 
   |\Phi_{m}^{(n-1)}\mkern3mu{C_m}\rangle
   \propto
   \left\{
      \begingroup\SmallColSep
      \begin{array}{l l}
         |\mkern1.5mu{0}\rangle
         |\mkern1.5mu{c_1}\rangle
         &
         \mkern3mu
         ,
         \mkern12mu
         a_{n.n}^{(n-1)}
         \mkern-5mu
         =
         0
         \\[5pt]
         |\mkern1.5mu{0}\rangle
         |\mkern1.5mu{c_1}\rangle
         +
         |\mkern1.5mu{a_{n.n}^{(n-1)}}\rangle
         |\mkern1.5mu{c_2}\rangle
         &
         \mkern3mu
         ,
         \mkern12mu
         a_{n.n}^{(n-1)}
         \mkern-5mu
         \neq
         0
      \end{array}
      \endgroup
   \right.
   \;\;\;\;    .
\end{equation}
\smallskip

\noindent After utilizing a measurement gate that collapses a quantum superposition onto one of its terms, one finds that the probability to get the zero outcome of the computation, denoted $\mathrm{Pr}(0)$, for each quantum processor correlates with the truth value of the statement $y$ such that\smallskip

\begin{equation}  
   \mathrm{Pr}(0)
   =
   \left\{
      \begingroup\SmallColSep
      \begin{array}{r l}
         1
         &
         \mkern3mu
         ,
         \mkern12mu
         {[\mkern-3.3mu[y]\mkern-3.3mu]}_v
         =
         1
         \\[5pt]
         0.5
         &
         \mkern3mu
         ,
         \mkern12mu
         {[\mkern-3.3mu[y]\mkern-3.3mu]}_v
         =
         0
      \end{array}
      \endgroup   
   \right.
   \;\;\;\;    .
\end{equation}
\smallskip

\noindent To guarantee that the above difference in $\mathrm{Pr}(0)$ will be detectable over errors, the QPRAM must have at least 3 processors. To be sure, let the null hypothesis, denoted $H_0$, be that the probability to get the zero outcome is 0.5. If $\mathrm{Pr}(0)$ is in fact 1, then the number of processors $q$, which gives 90\% power to reject $H_0$ (at the significance level of 0.10), will be three \cite{Fleiss}.\\

\noindent As a result, the cost of the quantum parallel computation required to check the solvability of the linear system $\mathbf{KX}\mkern-3mu=\mkern-2mu\mathbf{\Psi}$ is expected to be linear:\smallskip

\begin{equation} \label{QCOST} 
   \left[\mkern2mu{C_q}\mkern2mu\right]_y
   =
   O(n)
   \;\;\;\;     .
\end{equation}
\smallskip

\noindent This yields the outcome\smallskip

\begin{equation}  
   \left[\mkern2mu{E_q}\mkern2mu\right]_y
   =
   \frac{\left[\mkern2mu{W}\mkern2mu\right]_y}{\left[\mkern2mu{C_q}\mkern2mu\right]_y}
   =
   O(n^2)
   \;\;\;\;      
\end{equation}
\smallskip

\noindent demonstrating that the efficiency of quantum parallel computation can be much greater than 1.\\

\noindent More importantly, the relation (\ref{QCOST}) causes the discrimination in favor of qubits to be physically irrelevant:\smallskip

\begin{equation}  
   n
   \ge
   2
   \mkern-2mu
   :
   \mkern8mu
   \left[\mkern2mu{C_q}\mkern2mu\right]_y
   =
   \left[\mkern2mu{C}\mkern2mu\right]_x
   \;\;\;\;    .
\end{equation}
\smallskip

\section{Concluding remarks}  

\noindent If experimental atomic propositions relating to quantum systems are valuated within the ambit of a bivalent semantics, then, inevitably, the problem of an extra object emerges: Given three different objects, that is, three ordered pairs $(1,0)$, $(0,1)$ and $(0,0)$, but only two categories to put them into, i.e., the truth values of true, 1, and false, 0, what is the image of the pair $(0,0)$ under the Boolean function $b{\mkern-3mu}:{\mkern2mu}\mathbb{B}_2\times\mathbb{B}_2\to\mathbb{B}_2{\mkern2mu}$?\\

\noindent To resolve this problem, one can apply either of two workarounds.\\

\noindent Within Birkhoff and von Neumann's workaround, one assumes that the Boolean function $b$ is non-injective surjective and so $b(0,0)$ is equal to $b(0,1)$. The alternative way of resolving the problem of an extra object is to treat $b$ as partial Boolean function, thus allowing the pair $(0,0)$ to have no association with the elements of $\mathbb{B}_2$, i.e., to be a truth-value gap. Unlike Birkhoff and von Neumann’s workaround, the latter does not bring about the failure of the distributive law of propositional logic and dispersion-free probabilities for qubits. But despite this, it discriminates in favor of qubits.\\

\noindent Indeed, the workaround based on the assumption of a partial Boolean function gives rise to the fact that regarding experimental atomic propositions about many-level quantum systems, the number of primitive operations needed to assign the value of false is much greater than that of true. Whereas, for experimental atomic propositions relating to qubits, the assignment of either truth value requires the same number of those operations.\\

\noindent This discrimination, however, could be immaterial if there were to be a model of computation which permits execution of many primitive operations at once, so that the amount of time taken to solve a given computational problem can be much less than the number of primitive operations necessary to solve it sequentially.\\

\noindent As the present paper has demonstrated, such a model – which can be called a QPRAM, or quantum parallel random-access machine – is plausible.\\

\bibliographystyle{References}
\bibliography{Discrimination}

\end{document}